\documentclass{article}

\usepackage{PRIMEarxiv}

\usepackage[utf8]{inputenc} 
\usepackage[T1]{fontenc}    
\usepackage{hyperref}       
\usepackage{url}            
\usepackage{booktabs}       
\usepackage{amsfonts}       
\usepackage{nicefrac}       
\usepackage{microtype}      
\usepackage{fancyhdr}       
\usepackage{graphicx}       
\usepackage{multirow}
\graphicspath{{media/}}     
\usepackage{xcolor}
\usepackage{amsmath}
\usepackage{adjustbox}
\usepackage{float}

\usepackage{caption}
\usepackage{subcaption}

\usepackage{manyfoot}

\usepackage{lipsum}

\DeclareRobustCommand{\rchi}{{\mathpalette\irchi\relax}}
\newcommand{\irchi}[2]{\raisebox{\depth}{$#1\chi$}}
\newcommand{\rchisq}{$\rchi^2$}

\newcommand{\metric}{\psi}
\newcommand{\Metric}{\Psi}
\newcommand{\metricvalue}{m}

\newcommand{\stdmetric}{\hat{\metric}}
\newcommand{\metricos}{\hat{\metric}}

\newcommand{\metricvalueos}{\hat{\metricvalue}}
\newcommand{\Metricvalueos}{\hat{\metricvalue}}
\newcommand{\mmc}{\mathit{MMC}}
\newcommand{\samplefunction}{\Theta}
\newcommand{\sampler}{\theta}
\newcommand{\sshift}{\zeta}

\newcommand{\sscale}{\eta}

\newcommand{\sweight}{\alpha}
\newcommand{\Sweight}{\mathcal{A}}
\newcommand{\normfunc}{\Gamma}
\newcommand{\detwindow}{\omega}
\newcommand{\dwset}{\Omega}
\newcommand{\exam}{I}
\newcommand{\window}{l}
\newcommand{\stride}{s}

\newcommand{\coderepourl}{\url{https://github.com/microsoft/MedImaging-ModelDriftMonitoring}}

\newcommand{\footnotesept}{\hspace{.4em}\textsuperscript{,}}

\title{CheXstray: Real-time Multi-Modal Data Concordance for Drift Detection in Medical Imaging AI
}

\author{
  Arjun Soin
  \footnote{Joint first and corresponding authors}\footnotesept%
  \footnote{Microsoft Health and Life Sciences (HLS), Redmond, WA}\footnotesept%
  \footnote{Stanford Center for Artificial Intelligence in Medicine and Imaging (AIMI), Palo Alto, CA}%
  \\
  \texttt{asoin@stanford.edu} \\
   \And
  Jameson Merkow\footnotemark[1]\footnotesept\footnotemark[2] \\
  \texttt{jameson.merkow@microsoft.com} \\
  \And
  Jin Long\footnotemark[3] \\
  \texttt{jinlong@stanford.edu} \\
 \And
  Joseph Paul Cohen\footnotemark[3] \\
  \texttt{joseph@josephpcohen.com} \\
    \And
  Smitha Saligrama\footnotemark[2] \\
  \texttt{smitha.saligrama@microsoft.com} \\
     \And
  Stephen Kaiser\footnotemark[2] \\
  \texttt{steve.kaiser@microsoft.com} \\
    \And
  Steven Borg\footnotemark[2] \\
  \texttt{steven.borg@microsoft.com} \\
  \And
  Ivan Tarapov\footnotemark[2]\footnotesept\footnote{Joint senior authors} \\
  \texttt{ivan.tarapov@microsoft.com} \\
         \And
  Matthew P Lungren\footnotemark[2]\footnotesept\footnotemark[3]\footnotesept\footnotemark[4] \\
  \texttt{mlungren@stanford.edu} \\
}

\begin{document}
\maketitle
\begin{abstract}
\textbf{Background/Motivation}: Rapidly expanding Clinical AI applications worldwide have the potential to impact to all areas of medical practice. Medical imaging applications constitute a vast majority of approved clinical AI applications with more than 150 to date. Though healthcare systems are eager to adopt AI solutions after initial successes in model validation and deployment for clinical workflows, fundamental questions remain: \textit{Is the model still working as expected?, What is causing the change?, Is it time to intervene?} While infrastructure and local validation barriers are addressed during implementation, the larger problem looms regarding how to monitor input data and track model performance in real-time production systems. Real-world healthcare data environments are dynamic which can disrupt AI model performance, referred to as "drift". For example, medical equipment updates, new imaging workflows, changes in patient and disease characteristics, and more, can change over time and cause drift. While there are dozens of tools for monitoring drift, all lack functionality for image-based deep learning model architectures, cannot use healthcare medical imaging data, and require contemporaneous ground truth which is rarely available in healthcare settings. As a result, healthcare systems are left without insight for ongoing medical imaging model performance which poses serious safety concerns for healthcare system leadership, government regulators, and patients.

\textbf{Methods}: In this study, we utilize the CheXpert and PadChest open datasets to build, design, and test a medical imaging AI drift monitoring workflow that can track data and model drift in the absence of contemporaneous ground truth for a chest X-ray use case. Ten pathology labels were chosen and reconciled between the two datasets in order to fine-tune a densely connected convolutional neural network on PadChest after being pre-trained on frontal-only CheXpert images. The PadChest dataset and accompanying ground truth labels were chronologically arranged and split such that all fine-tuning and validation used only data between 2007-2013 while tests were performed on data between 2014-2017. To test the robustness of our approach, we simulate drift in a variety of experiments and observe measured performance and drift within a data stream. We measure performance using ground truth labels and compare our drift metric with actual model performance over time in the test dataset and identify correlative signals.  Drift evaluation used DICOM metadata, image appearance, and model output scores as input. Variational autoencoders (VAE) were used to capture an appearance feature encoding by using the underlying latent space representation.
For continuous features (e.g. age, latent encoding from VAE), Kolmogorov-Smirnov (K-S) tests are applied, and for categorical features (e.g. sex, view projection), \rchisq tests are used to measure the distribution shift.

\textbf{Results}:  In aggregate, we found agreement between our proposed multi-modal data concordance metric and medical imaging AI model performance metrics. Through experimentation, we demonstrate a strong proxy for ground truth performance (AUROC) using unsupervised distributional shifts in relevant DICOM metadata tags, predicted probabilities, and VAE latent representation. 
This comprehensive approach to unsupervised drift detection was found to correlate with supervised drift detection approaches leveraging ground truth labeling data.

\textbf{Conclusion}: We propose methodologies to achieve real-time drift monitoring metrics in the absence of contemporaneous ground truth in a medical imaging AI model and demonstrate the robustness of our approach with a chest X-ray use case. Our key contributions include (1) proof-of-concept for medical imaging drift detection approaches including use of variational autoencoders and medical imaging data specific statistical methods (2) a methodology for measuring and unifying drift metrics across patient demographics, imaging metadata and pixel based statistics (3) new insights into the unique challenges and proposed solutions for observing medical imaging AI models in production (4) creation of open-source tools\footnotemark[1] 
that leverage existing open source medical imaging datasets, enabling others to easily use our tools. 
This work has important implications for addressing the translation gap related to continuous medical imaging AI model monitoring in dynamic healthcare environments.
\parbox{.9\textwidth}{\footnotesize \vspace{.8em} \flushleft \footnotemark[1]Code available at: \coderepourl}
\end{abstract}

\section{Introduction}

Artificial intelligence (AI) applications in medical imaging have expanded substantially over the past 5 years \cite{van2021artificial}.  
The growth is evident by both the rising volume of academic publications and the acceleration of commercial approvals for these applications in clinical practice \cite{west2019global, benjamens2020state, van2021artificial, mehrizi2021applications, tadavarthi2020state}. Alongside this trend of new discovery and market-ready products\footnote{\href{https://www.fda.gov/medical-devices/software-medical-device-samd/artificial-intelligence-and-machine-learning-aiml-enabled-medical-devices}{FDA | Artificial Intelligence and Machine Learning (AI/ML) Enabled Medical Devices}}, clinicians are increasingly eager to adopt AI solutions into their practice \cite{tariq2020current}.  However, to date clinical translation has been disproportionately limited. The reasons behind the translational gap in real-world clinical practice are multi-factorial, partially explained by technical and infrastructure hurdles, lack of IT resources, and no clear data-driven clinical utility analyses. Many of these barriers to adoption are being addressed with existing or emerging solutions \cite{wiggins2021imaging, mahajan2020algorithmic, eche2021toward, dikici2020integrating}.  Yet even with successful site specific model validation and successful integration/deployment in clinical workflows, a fundamental problem remains: what happens after the AI model goes into production? 

Of particular interest in these production systems is how does model performance change over the life cycle of an AI model.
Traditional performance drift detection requires monitoring a metric of interest (such as AUROC, F1 score, PR scores) then alerting when that metric falls below a specified value, allowing administrators to investigate, perhaps triggering an adjustment or retaining of the model. Clearly translating medical imaging AI safely and effectively requires a real-time understanding of performance to address critical questions that remain unanswered in healthcare AI monitoring.
The lack of visibility and the inability to guard against performance drift remains a critical barrier to widespread adoption of AI solutions in healthcare \cite{Finlayson2021TheCA}.  The current lack of answers to these questions in the field demonstrate the unrealistic expectation that input data and model performance will remain static indefinitely which runs counter to decades of machine learning operations research, as outlined by extensive experience in AI model deployment for other verticals \cite{sculley2014machine, klaise2020monitoring}. Identifying a solution for real-time model monitoring in production for clinical workflows is crucial, and includes detecting both out-of-distribution data as well as data drift using statistical techniques. Most performance monitoring solutions in production environments require systematic access to contemporaneous or near real-time ground truth data to inform metrics \cite{rabanser2018failing}. But in healthcare, ground truth data is seldom, if ever, available in real-time, particularly for medical imaging. Further, existing model monitoring solutions are designed to leverage structured tabular data, and no solution currently exists for imaging data. The challenge we face in the medical imaging AI model monitoring task, then, is to derive a systematic approach to real-time clinical AI model performance monitoring for medical imaging data (pixel and non-pixel data) without contemporaneous ground truth labels. To tackle this critical issue, we propose a system that relies on statistics of input data, deep-learning based pixel data representations, and output predictions coupled with a novel multi-modal integration solution to allow real-time monitoring that can alert when data has drifted which may adversely affect model performance; a solution which \textit{to date has never been described for medical imaging models}.  

Monitoring of machine learning models in production is a distinct domain, that lies between traditional software systems and quality outcome management. It requires appropriate practices, strategies, and tools \cite{sculley2014machine}.
These challenges are exacerbated in medical imaging by the lack tools for monitoring pixel-based AI models across the field. Further, medical imaging data is often accompanied by various metadata regarding the patient demographics, device model and manufacturer, patient position, image projection, and a number of device settings all of which may lead to unexpected results from AI systems.  The predictive performance of medical imaging AI models can degrade in drift scenarios such as changing patient populations, disease prevalence, acquisition protocols, new imaging software equipment or updates, new clinics, and many more. Furthermore, it is insufficient to simply apply methodologies to measure changes in these statistics individually for practical notification and intervention as data points per day may number in the hundreds or thousands and while changes in some are critical, many are superfluous. Thus, unifying the multitudes of individual metrics appropriately is vital to monitoring drift for healthcare AI.

In this manuscript, we explore a data-driven approach to a system for real-time AI model monitoring in a medical imaging environment. We demonstrate a critical use case of providing drift metrics that correlate strongly to changes in model performance (based on ground truth metrics) but do so without the needs for contemporaneous ground truth. The key contributions of the presented work include (1) multifaceted medical imaging drift detection approach including use of variational autoencoders and statistical methods (2) a methodology for measuring and unifying drift metrics across patient demographics, imaging metadata and pixel based statistics (3) new insights into the unique challenges and proposed solutions for medical imaging AI model in production (4) creation of open source tools, demonstrated on existing public datasets, that allow the research community to build and validate their own custom monitoring systems.

\begin{figure}
  \centering
  \includegraphics[width=1\columnwidth]{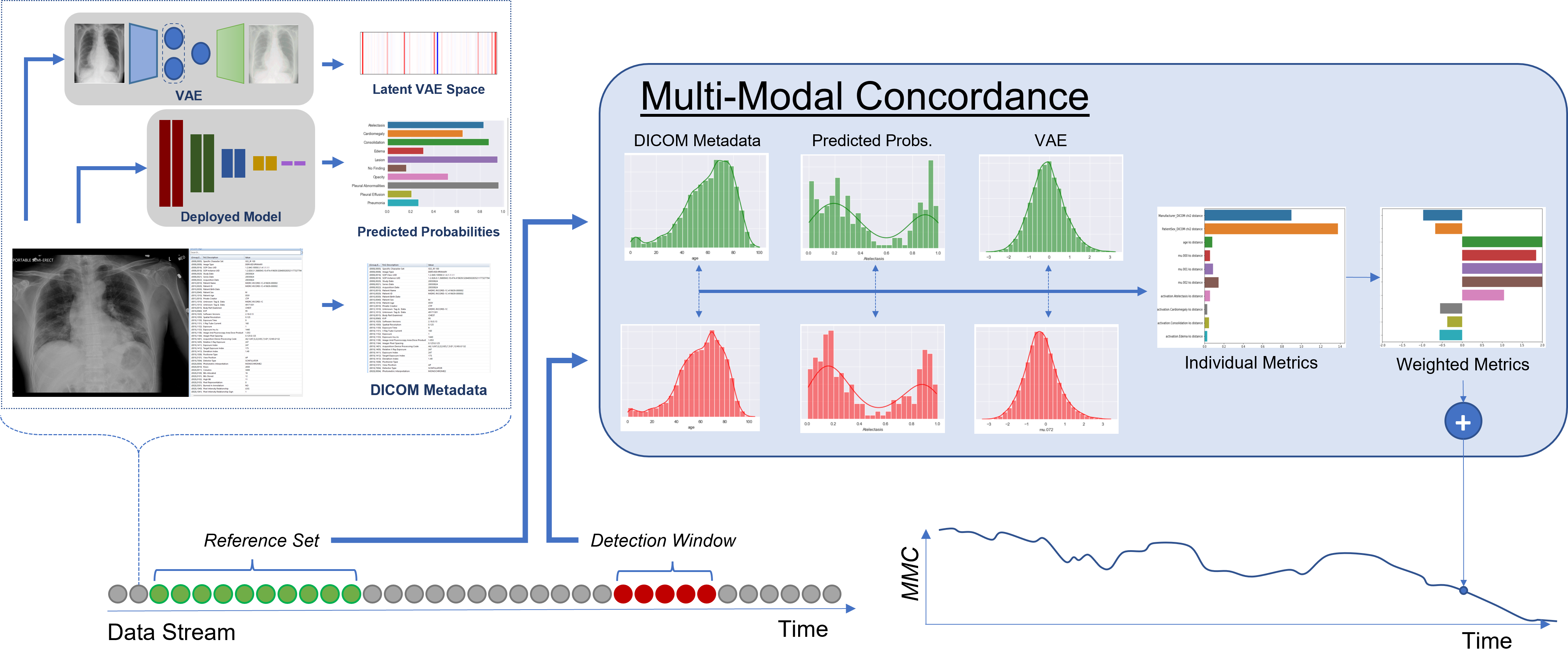}
  \caption{Overview of our Multi-modal concordance algorithm. %
  From each object in a data stream of x-ray exams, we extract DICOM metadata, model predicted probabilities and a latent representation produced by a variational autoencoder.  We collect these values from exams in a reference set, and compare distributions of extracted data to a detection window to produce similarity for each component.  We then standardize and weight these measures to combine them into a single value representative of the total concordance between the reference data and the detection window.  This provides with a simple metric capable of detecting data drift from the reference.
  }
  \label{fig:system}
\end{figure}

\section{Methods}
\label{sec:headings}
In this section, we focus on outlining the various building blocks of our approach to medical imaging AI drift detection, cover data specifications, AI model, drift detection concepts and foundational metrics. 
\subsection{Data}

Our experimentation utilizes two publicly available datasets to test a medical imaging AI drift workflow: CheXpert \cite{chexert} and PadChest \cite{padchest}.  CheXpert is a large dataset of chest X-rays and competition for automated chest X-ray interpretation that includes uncertainty labels as well as a radiologist-labeled reference standard evaluation sets. CheXpert contains 224,316 chest radiographs of 65,240 patients who underwent an examination at Stanford University Medical Center between October 2002 and July 2017 and includes both inpatient and outpatient medical scans. 

The PadChest dataset is a large, labeled chest X-ray dataset containing 160K high resolution images with their corresponding labeled reports. Its detailed description and labeling methods are described in \cite{padchest}. PadChest includes more than 160,000 images obtained from 67,000 patients that were interpreted and reported by radiologists at San Juan Hospital (Spain) from 2009 to 2017, covering six different position views, and including additional information on image acquisition and patient demography. Both human annotation and Natural Language Processing (NLP) annotation was used to  obtain PadChest’s labels. 
Labels include 19 differential diagnoses, 103 anatomic locations and 179 different radiological findings which were mapped onto the NLM standard Unified Medical Language System (UMLS) using controlled biomedical vocabulary unique identifiers (CUIs). These were further organized into semantic hierarchical concept trees. Unlike many other datasets, the chronology of the scans was not removed during de-identification, making it uniquely useful in model and data drift experiments. 

\subsubsection{Data Framework \& Chronology}
\label{sec:dataframework}
Typically, training a new clinical model will leverage a pre-trained model which is fine-tuned on a new task and class biases of a different clinical setting. To mimic this process, we started with an available model, pretrained on CheXpert then trained and validated it on PadChest data.
Fine-tuning a model on PadChest data from a model pretained on CheXpert, requires unifying their labels sets.  PadChest original has an extensive label set from which we merged relevant labels into a common label set of ten: Atelectasis, Cardiomegaly, Consolidation, Edema, Lesion, No Finding, Opacity, Pleural Abnormalities, Pleural Effusion and Pneumonia (See Table \ref{tab:labels} for details).

We split PadChest in training, validation and test sets based on their exam dates, allowing us to produce simulated data streams from the data. We used {2013-01-01} and {2014-01-01} to partition the data into training, validation and test sets.
Our training data spans {2007-05-03} to {2012-12-31} (12-29 to 12-31 contain frontal samples), validation data 2013-01-01 to 2013-12-31 and finally our test data starts on {2014-01-01} and continues to the end of the dataset ({2017-11-17}). 
Note that in early 2014 the method for which image were labeled began to change; where studies were previous labeled using NLP, they were now manually labeled by expert radiologists.
There are also gaps in the data where the number of exams per day drops significantly.
Since most AI models have a life cycle of about one year, we conduct our experiments  within the first year of the test set (from {2014-01-1 to 2014-12-31}). 
For a CheXpert pathology label-wise training data distribution breakdown, refer to \cite{chexert}.
Table \ref{tab:padchest-demo} contains the resulting post label mapping distribution and date cutoffs between domains (training, val, test).

\begin{table}[h]
\centering
\caption{PadChest Condensed Labels}
\label{tab:padchest-demo}
\begin{tabular}{@{} llll @{}}
\toprule
& Training & Validation & Test\\
\midrule
Start Date & 2007-05-03 &2013-01-01&2014-01-01\\ 
End Date & 2012-12-28 &2013-12-31&2014-12-31\\
Total Images & 63,699 &15,267&11,509\\ \midrule
&\multicolumn{3}{l}{\textbf{Pathology Counts}} \\ \midrule
Atelectasis & 4,516 &1,121&954 \\
Cardiomegaly & 5,611 & 1357 &935 \\
Consolidation & 1,161 &225&106 \\
Edema & 26 &9&17 \\
Lesion & 1,950 &390&294 \\
No Finding & 21,112 &4,634&3,525 \\
Opacity & 11,604 &2,577&2,018 \\
Pleural Abnormalities & 6,875 &1,708& 1,272 \\
Pleural Effusion & 4,365 &1,026& 710 \\
Pneumonia & 3,287 &584&379 \\
\bottomrule
\end{tabular}
\end{table}

\subsection{Deep Learning Model and Baseline Performance}
For our classifier model, we used a Densely Connected Convolutional Neural Network \cite{Huang2017densenet} with 121 layers as originally implemented in Pytorch for the CheXpert competition.  Starting from ImageNet \cite{imagenet_cvpr09} weights, we first pretrain the model on frontal-only CheXpert training data (N = 191,000), using a \textit{U-Ones} scheme (uncertain labels considered positive) as outlined in \cite{chexert}, We selected the model that yielded the best performance on the 10 listed pathologies in section \ref{sec:dataframework} and then retrain only the final classifier layers on PadChest frontal-only\footnote{View Positions of PA, AP or AP\_horizontal projections as frontal} training data.
After PadChest training completes, we deploy the model in a psuedo-clinical setting and pass the entirety of PadChest through the system in a sequential, date-wise fashion, recording both raw scores and activations from the system for each sample. Using these activations, we are able to measure performance benchmarks for any time window within the dataset. The performance benchmarks for each stage of model development are found in Table \ref{tab:chexpert}.

Note that while transfer learning from ImageNet for downstream medical imaging tasks is the current standard in Deep Learning, we fine-tune the model on two large frontal X-ray datasets to unify the pathology labels, allowing us to accurately explore different drift scenarios in a pseudo-clinical deployment setting. 

\begin{table}[ht]
\caption{Model Performance (AUROC) on CheXpert and PadChest, frontal-only}
\label{tab:chexpert}
\centering
\begin{adjustbox}{width=1\textwidth}
\begin{tabular}{@{} rccccccccccc @{}}
\toprule[\heavyrulewidth]
 \multirow{3}{*}[-0.25em]{\rotatebox[origin=c]{90}{CheXpert}}& & Atelectasis & Cardiomegaly & Consolidation & Edema & Lesion & No Finding & Opacity & Pleural Abnormalities & Pleural Effusion & Pneumonia \\ \cmidrule[\heavyrulewidth](l){3-12}
&Training                      & 0.872   & 0.964   & 0.908   & 0.945  & 0.966  & 0.979  & 0.882  & 0.967  & 0.959  & 0.916   \\ \cmidrule[\heavyrulewidth](l){2-12}
&Validation                    & 0.632  & 0.794   & 0.817   & 0.857  & 0.783   & 0.85  & 0.872  & 0.963  & 0.888  & 0.684  \\ \midrule[\heavyrulewidth]
\multirow{3}{*}[-0.25em]{\rotatebox[origin=c]{90}{PadChest}}& & \multicolumn{10}{c}{} \\ \cmidrule[\heavyrulewidth](l){2-12}
&Training   &       0.831 &        0.933 &         0.906 &  0.920 &  0.804 &      0.877 &   0.885 &                 0.926 &            0.959 &     0.870 \\ \cmidrule[\heavyrulewidth](l){2-12}
&Validation &       0.793 &        0.922 &         0.892 &  0.915 &  0.776 &      0.862 &   0.867 &                 0.933 &            0.966 &     0.878 \\ \bottomrule[\heavyrulewidth]
\end{tabular}
\end{adjustbox}
\end{table}

\subsection{Data Stream Drift}
When objects in a dataset include timestamps, it is referred to as a data stream. The underlying statistics and properties of a data stream are subject to change over time, which gives rise to drift. Broadly speaking, drift falls into one of two categories: Covariate shift (sometimes called input or feature drift) and concept shift. 

Covariate shift \cite{Moreno-Torres2012datasetshift} is defined as a deviation within the input variable of a data stream.  Covariate shift is common in medical imaging, examples include changes to imaging protocols, imaging software or equipment updates, and changing patient demographics.
After a covariate shift, a deployed model may be operating in an untested or poorly validated environment wherein 
performance degradation becomes an obvious concern \cite{de2022guidelines}. 
When a significant time gap exists between contemporary data and model deployment, the likelihood of drift and consequently, classification errors increases \cite{Finlayson2021TheCA}. 
For healthcare AI systems, drift associated errors can cause unwarranted harm to patients.
Ideally, we would like to continuously monitor any clinically deployed AI system, and refresh the algorithm with new training data upon observing any significant performance degradation.

Whereas the covariate shift refers to changes in input data $x$, concept shift occurs when the relationship between input data $x$ and output variable $y$ changes. Modern AI systems are built upon \textit{stationarity} - the idea that the characteristics of a target class remain static \cite{conceptdrift}. 
This assumption allows models to be trained to identify those characteristics then predict the presence of that class in unseen data. 
This assumption is not always valid, particularly when the target class can be influenced by outside factors.
Take for instance, the impact of COVID-19 on an automated chest X-ray interpretation model trained pre-COVID. A model designed to predict mortality for COVID patients using chest radiograph images may work on a dataset taken from the height of the pandemic but as treatments advance, disease prevalence shifts, the mortality outcome based on imaging features may no longer be sufficient for accurate prediction.
In this work, we assume that concept is fixed and our experiments concentrate on detecting changes related to covariate shift.

\subsection{Concordance Measurement}

In our approach, instead of highlighting differences in the data stream, we monitor the similarity or concordance of the data stream with respect to a reference dataset; when the concordance metric decreases the degree to which the data has drifted has increased.
Specifically, our method measures the concordance between `reference' set and second set we refer to as a `detection' window. The `reference' set comprises a gold standard collection of samples with known characteristics (and model performance) with which we wish to stay in concordance with. To measure this concordance on a detection window, we calculate a number of individual metrics that compare statistics between the two samples.  We calculate these metrics sequentially on a data stream providing concordance metrics over-time. More formally, we define a reference set, $\detwindow_R$ that is a collection of individual exams ($\exam$), $\detwindow_R = \{\exam_0, \exam_1, \cdots, \exam_{K-1}, \exam_K\}$. Using this sample, we are able to measure concordance of a detection window at time $t$ ($\detwindow_t$) by applying our collection of metric functions, $\Metric = \{\metric_1, \metric_2, \cdots, \metric_{N-1}, \metric_N\}$, each metric compares a subset of statistics between the two samples.


For metric functions, we chose two statistical tests, one for continuous real-valued features and another for discrete (categorical) valued features. For continuous features (e.g. age, $z$ from VAE), a Kolmogorov-Smirnov (K-S) test is applied which measures distribution shift from the reference window. 
The K-S test is a non-parametric test used to measure distribution shift of a continuous variables from a reference sample. As a non-parametric test, the K-S test compares samples without assuming a specific distribution of a variable making it an efficient and effective way to distinguish the distribution change from one time to another \cite{kstest}. While we used these metrics for our experimentation, our framework is extensible and modular, built specifically to allow additional metrics to be seamlessly added.

For categorical features (e.g. sex, projection), the \rchisq (chi-square) goodness of fit test is used to compare observed frequencies in data and compares them to expected values. Another non-parametric test, \rchisq goodness of fit test, calculates if an input sample with observed frequencies is likely to be obtained from the frequencies observed in the reference set\cite{Pearson1900}.

Both of these tests provide both a statistical measure of the similarity between the two distribution as well as a p-value that provides a likelihood that the null hypothesis is accepted or rejected. We found that the p-value gives noisy results and to be an inconsistent measure of similarity between two distributions. On the other hand, the test statistics directly compare the two distributions and provide a softer and more consistent metric for measuring similarity. For these reasons, our experiments concentrate on the test statistics for measuring concordance and ignore the p-values for both tests.

\subsection{Performance Measurement}

To evaluate performance of our model, we calculate AUROC, which serves as a discrimination measure to demonstrate how well the model can classify patients in two groups: those with and those without a given pathology of interest \cite{auroc}. The AUROC is the integral of the receiver operating characteristic curve which measures the trade-off between true positive rate (TPR) and false positive rate (FPR) at different decision thresholds. A test with no better performance than chance has an AUROC of 0.5, while a test with perfect accuracy would have an AUROC of 1.0. Accordingly, an AUROC = 0.90 indicates that 90\% of the time we draw a test X-ray from the disease group and non-disease group, the predicted score from the disease group will be greater.

Since monitoring AUROC over production time-frame can provide clear-cut evidence of a model drifting, it is a metric that physicians, hospital systems and AI clients would ideally track real-time. However, this also requires real-time, domain expert-labeled ground truth labels (a cost-prohibitive and unfeasible ask even for the most dynamic healthcare systems). AUROC serves another critical purpose for this work, giving insight into model performance-based drift analysis in contrast with other statistical metrics that aim to do so without ground truth. 

\section{Multi-Modal Concordance Framework}
\label{sec:others}

We measure data concordance by capturing metrics from three sources: 1) DICOM metadata, 2) image appearance data, and 3) model response data, then unify those metrics into a single multi-modal concordance metric which we refer to as Multi-Modal Concordance ($\mmc$).
Our multi-modal metric is comprised of a diverse set of signals that cover a variety of areas where drift can occur. 
DICOM metadata contains information on the origin and construction of the image as well as patient demographics; changes in these variables may be indicative of feature drift. 
In addition to image source information, our approach utilizes appearance-based features from a Variational Autoencoder (VAE) to directly monitor visual feature drift.
Lastly, we incorporate model responses which could indicate if other changes have occurred that directly affect model output statistics including covariate shift or prior probability shift.

 \subsection{DICOM Metadata \& Statistical Distribution Shift}
 
 DICOM (Digital Imaging and Communications in Medicine) is the international standard to transmit, store, retrieve, print, process, and display medical imaging information \cite{mildenberger2002introduction}. Data available in the DICOM format is produced by the imaging device. A DICOM file consists of a header and image pixel intensity data packed into a single file. An image header represents embedded metadata. This metadata includes demographic information such as patient sex and date of birth, as well as a record of imaging attributes that govern how the image is captured, stored and transmitted. All of these attributes can effect an AI system's responses. We use these DICOM variables to characterize data shifts given that changes to these attributes can be indicative of changes to imaging features and patient population makeup. For metadata, clinical and imaging protocols extracted from the PadChest dataset are analyzed to generate the drift metrics in this study and fall into 3 categories: They are 1) patient demographics features including age and sex, 2) image formation metadata, i.e. X-ray scan information including view position, device manufacturer, frontal projection (Y/N), X-ray tube current, X-ray exposure and relative exposure, 3) image storage information such as pixel representation, spatial resolution, bits stored, window width, and pixel aspect ratio.


\subsection{Appearance-Based Shift}
 

Medical imaging data is complex, and a medical imaging data stream can shift without any accompanying metadata changes. Change in imaging hardware, disease presentation or patient demographic that are not captured in the DICOM metadata may be invisible to a human eye, but noticeable to the sensitive ML model. After such a shift, features originally present in the training and validation data may morph or disappear altogether, impacting model performance. It is crucial to be able to capture these changes when monitoring a live data stream. Recent work for high-dimensional data drift detection proposes combining dimensionality reduction (eg. PCA, autoencoders etc.) with two-sample hypothesis testing \cite{rabanser2018failing}. We leverage a Variational Autoencoder (VAE) to generate an encoded representation of each image upon which we apply statistical metrics to detect drift. Rather than measuring change by image reconstruction loss which is a scalar quantity, we utilize the latent space encoding generated by the VAE which gives us a feature-rich representation with compressed yet relevant information about the input images.
This a encoded representation can more easily be checked for distributional shifts than the pixel representation \cite{Zenati2018, Cao2020-OoD-Benchmark, Shafaei2018} and is more descriptive than a scalar reconstruction value.
Using this feature rich latent space allows a much more fine-grained (and often explainable) analysis.

Generic autoencoders compress input data into an encoded representation, then reconstruct the original data using only the latent values.  
Based on this concept, variational autoencoders (VAE) assume that the data follows some underlying parametric probability distribution (typically a multi-variate Gaussian) and attempts to model the parameters of this distribution which become the image's latent representation.
A by-product of this process is that the input is not simply compressed, but is encoded into a probabilistic latent space where inputs where similar features have similar latent representations.\cite{vae2019}. 
We leverage this fact and encode input images into this descriptive latent representation, then build a statistical model of it, allowing us to compare our reference data with a detection window. 
\begin{figure}
  \centering
  \includegraphics[width=.95\columnwidth]{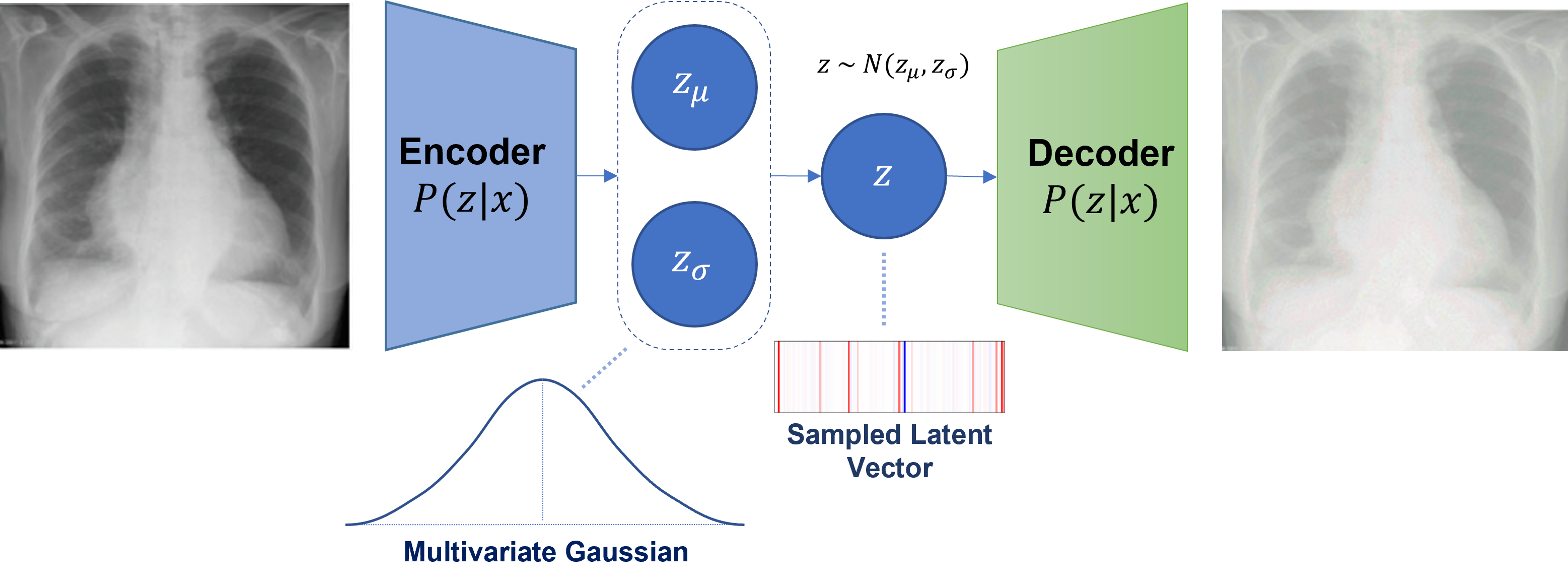}
  \caption{Variational Autoencoder (VAE)}
  \label{fig:fig3}
\end{figure}

We trained our variational autoencoder (VAE) from scratch on the PadChest training data including both frontal and lateral images.
During test time, we feed all input images within the detection window through the encoding portion of the VAE and capture the resulting $z$ parameters of the probability distribution. These parameters are used to establish statistical similarity of new data to our reference dataset. 

\subsection{Model Response Drift}
The goal of monitoring live medical data streams is to measure the consistency of a data stream over the life cycle of a deployed model. If this data stream begins to drift, and model performance begins to degrade, it stands to reason that the model responses would also change. Thus, it is crucial to monitor the model responses for any such changes. By and large, model responses shift for two reasons: 1) the underlying class distribution has shifted (prior probability shift), or 2) the visual representations have shifted (covariate shift). Both of these conditions can dramatically affect performance, in particular the latter may have a significant impact on model performance, but would likely go unnoticed without a performance audit. Measuring these responses directly allows us to catch any such changes in the data stream.

In general, model response monitoring takes one of two forms: measuring soft predictions or hard predictions.  Soft predictions refer to the raw score or activation from the model, whereas hard predictions refer the the final label out from the system, typically after applying a threshold. 
In this work, we exclusively use soft predictions for monitoring. Soft predictions contain valuable information on the relative certainty of our predictions.  Model responses may incrementally shift as rare or more subtle signs of pathology increase in frequency, a change that would remain hidden by hard predictions until the responses pass some threshold.  By directly measuring shifts in soft responses, we are able to detect these subtle changes.

\subsection{Metric Unification}

In this section, we discuss how our approach unifies multiple metrics across a variety of multi-modal inputs. Specifically, we discuss our sampling methodology for creating detection windows as well as our strategy to standardize and aggregate individual metrics into a single multi-modal data concordance metric.

\subsubsection{Sampling Methodologies}

\textbf{Rolling Detection Window}:
To construct our detection windows, we use a sliding window technique. Sliding window sampling functions have parameters for window length ($\window$) and stride ($\stride$). The window length determines the size of each window and the stride denotes the spacing between neighboring samples.  We use temporal-based stride and window length values to collect all exams within a time-window, looking backwards from the indexed date.
For example, if the window length is 30 days, then sample for December 31st would include all exams from December 1st through December 31st.
Note that, our approach calculates multiple metric values, $\metricvalue_i$ from each detection window, $\detwindow$ using metric functions $\Metric_i(\detwindow) =  \metricvalue_i$.

\textbf{Over Sampling}:
Many distribution similarity metrics, including those those used in this work, are sensitive to sample size; even when two samples are drawn from the same distribution they may produce differing results simply due to sample size. Our method mitigates this issue by repeatedly calculating metrics on a fixed-size sample and calculating an average result. To do this, we use a bootstrap method that samples the detection window to draw $K$ samples, then calculate metrics on this bootstrapped sample.  We repeat this process $N$ times and average the results to obtain a final value.
More formally, computation of a given metric on a detection window is a  function $\samplefunction(\cdot)$ that uses the sample and another function, $\sampler^K$ to collect $K$ samples from a detection window $\detwindow$ to calculate metric $\metric_i$, which is done $N$ times then the results are aggregated:
\begin{equation}
    \samplefunction_{\metric}(\detwindow,N,K) = \metricos_i(\detwindow) = \metricvalueos_i = \frac{1}{N}\sum_0^N \metric_i\left(\sampler^K(\detwindow)\right)
\end{equation}
where $\sampler^K$ is a function that collects $K$ samples from $\detwindow$ with replacement.

\textbf{Detection Window Sets}:
A detection window set is a collection of detection windows, where each detection window is typically captured with a time-index.
If we denote a detection window taken at time $t$ as $\detwindow^t$, then we define some detection window set taken from time $a$ to time $b$ as $\dwset^{[a,b]} = \{\detwindow_a, \detwindow_{a+1}, \cdots, \detwindow_{b-1}, \detwindow_{b}\}$.
We can then collect metrics at each time step, resulting in a metric value set at multiple time steps: we define $\metricos_{i}(\detwindow^t) = \metricvalueos_i^t$ as an individual metric calculated at time $t$ from detection window $\detwindow^t$, we can then capture metric values $\Metricvalueos_i^{[a, b]} = \{\metricvalueos_i^a, \metricvalueos_i^{a+1}, \cdots, \metricvalueos_i^{b-1}, \metricvalueos_i^b\}$ as the collection of $\metricvalueos_i^t$ from time $a$ to time $b$. Consequently, we use $\Metricvalueos_i^{[a, b]}$ as an individual measure of drift over time.

\subsubsection{Metric Standardization and Aggregation}
Metrics outlined in the previous sections provide diverse inputs for monitoring drift, however, presenting a cohesive framework that bridges supervised and unsupervised data requires a holistic approach and metric unification.
There are three main challenges to metric unification:  1) fluctuation normalization, 2) scale standardization and 3) metric relevancy.
Without normalizing for acceptable fluctuation, it is impossible to differentiate changes that occur within normal operation and those that truly represent drift.
Furthermore, since each of these metrics are based on different types of tests comparing separate statistics, there is no guarantee that these values will reside on the same relative scale.  For example, a \rchisq test statistic has no upper bound, where as a two sample K-S test statistic lies between 0 and 1.
Merging metrics across non-standardized values may result in improper unification, where large values overpower smaller ones disregarding their relative importance.
We tackle fluctuation normalization and scale standardization by using a standardization function, $\normfunc$, which transforms all individual metrics into a numerical space with common upper and lower bounds.  In this work, we use a simple function that normalizes an input value $m$ with fixed values for scale and offset:
\begin{align}
\normfunc(m) = \frac{m-\sshift}{\sscale}
\end{align}
where scale and offset factors are represented by $\sscale$ and $\sshift$, respectively.

Next, we unify individual metric values across all standardized metrics through weighted aggregation using predefined weights, $\sweight_i$ for each metric.
Putting it all together, we calculate our multi-modal concordance metric, $\mmc$, on a detection window $\detwindow$ from $L$ metrics as follows:
\begin{equation}
    \mmc(\detwindow) = \sum_{i=1}^{L} \sweight_i \cdot \normfunc_i\bigg(\metricos_i(\detwindow)\bigg) = \sum_{i=1}^{L}\frac{\sweight_i}{\sscale_i}\bigg(\metricos_i(\detwindow)-\sshift_i\bigg)
    \label{eqn:mmc}
\end{equation}
where $\metricos_i(\detwindow)$ represents the $i$th metric calculated on detection window $\detwindow$, $\normfunc_i$ represents the standardization function, and $\sweight_i$ represents the weight used for the $i$th metric value. Calculating $\mmc$ on a time-indexed detection window set $\dwset^{[a,b]}$, we now have a robust multi-modal concordance measure capable of monitoring drift over the given time period from $a$ to $b$, $\mmc^{[a,b]}$.

A number of strategies exist to choose appropriate values for $\sscale$, $\sshift$ and $\sweight$; these strategies range from manual selection to fully automated functions.  Indeed, each of the metric weights, scales and offsets could be manually chosen using clinical heuristics.
In our experiments, we used automatic methods to calculate values for $\sscale$, $\sshift$ and $\sweight$.
Instead of manually choosing weights which may be time consuming, we propose an automatic method for obtaining scale and offset values, as well as metric weights.
First, we calculate values for $\sscale_i$ and $\sshift_i$ using a detection window set collected from the validation data.
Specifically, we first generate raw metric values using individual metric functions $\metric_i$, on all windows in a detection window set, calculate the means and standard deviations of each metric value across the detection window set and finally set each $\sshift_i$ and $\sscale_i$ to their corresponding mean and standard deviation.
Second, we obtained values for each $\sweight_i$ using a strategy that ties concordance directly to performance by leveraging the correlation between individual metrics, $\metric_i$ and performance on validation data.
Specifically, each weight, $\sweight_i$, is calculated using a detection window set $\dwset_{\sweight}$ as follows:
\begin{align}
    \sweight_i &= \big\lvert {corr}(\stdmetric_i(\dwset_{\sweight}), \rho(\dwset_{\sweight})) \big\rvert
\end{align}
where $\stdmetric_i(\dwset_{\sweight})$ and $\rho(\dwset_{\sweight})$ represent the standardized metrics and performance on some detection window set $\dwset_{\sweight}$. Selection of $\dwset_{\sweight}$ requires careful consideration. The detection window set $\dwset_r$ used to standardize metrics is often not suitable as it contains only high performing samples (by design).
We generate $\dwset_{\sweight}$ by adding poor performance samples into $\dwset_r$ though hard data-mining of our validation set. 

For experimentation, we also use a baseline method for calculating weights, where all weights are equal and we instead apply a simple average across standardized metric weights.
Throughout this work, we denote $\mmc$ calculated without any weights as $\mmc_0$ and the weighted counter part as $\mmc_w$.
Note that depending on the individual metrics used, the sign of $\sweight$ may need to be flipped to measure similarity rather than distance (as is the case for both statistical tests used in this work).

\section{Experimentation}

In this work, we aim to investigate the connections of various drift detection outcomes over the lifetime of a medical imaging model deployed in production. To observe these connections, we designed an experimental setup with an adaptive learning workflow backed by our open-source framework\footnote{\coderepourl} and stress-tested it on engineered drift scenarios that simulate different indicators for medical imaging AI drift in a production environment. Our framework has a modular design and can be used in a plug-and-play manner to test multiple input drift modalities and scenarios with include or new datasets.

Using this framework, we highlight two real-time drift scenarios by either generating an artificial data stream backed by real data or injecting data into a real data stream to induce drift, Thus, enabling us to retain genuine data properties while ensuring drift substantial enough to risk degrading the model.

\subsection{Experimental Setup}

In all of our experiments, we used the validation set for the reference data from which we wished to detect concordance/drift.  We also used the validation set to generate metric standardization and metric weights. 
We used data starting 2014-01-01 and ending on 2014-12-31 as our test set to simulate a production data stream.
Manipulation of this test data stream forms the basis for drift observations.
All experiments used a detection window stride ($\stride$) of 1 day and length ($\window$) of 30 days. For our sampling function $\samplefunction$, we set $K=2500$ and $N=20$. Note that, in some situations, the number of exams in a detection window can be extremely low, leading to anomalies in our metric calculations.  For this reason, we skip any detection windows that contain fewer than $150$ exams.
We used these same values to produce a reference detection window set, $\dwset_{r}$, which was used to calculate standardization factors $\sscale$ and $\sshift$.  Lastly, we generated an additional detection window set, $\dwset_{\sweight}$ to calculate metric weights, $\sweight_i \in \Sweight$. This detection window set was obtained by augmenting $\dwset_{r}$  with poor performing samples through hard data mining.

\subsection{Drift Scenario 1: Performance Degradation}
\label{sec:drift1}
\begin{figure}[hbt!]
  \centering
  \includegraphics[width=\columnwidth]{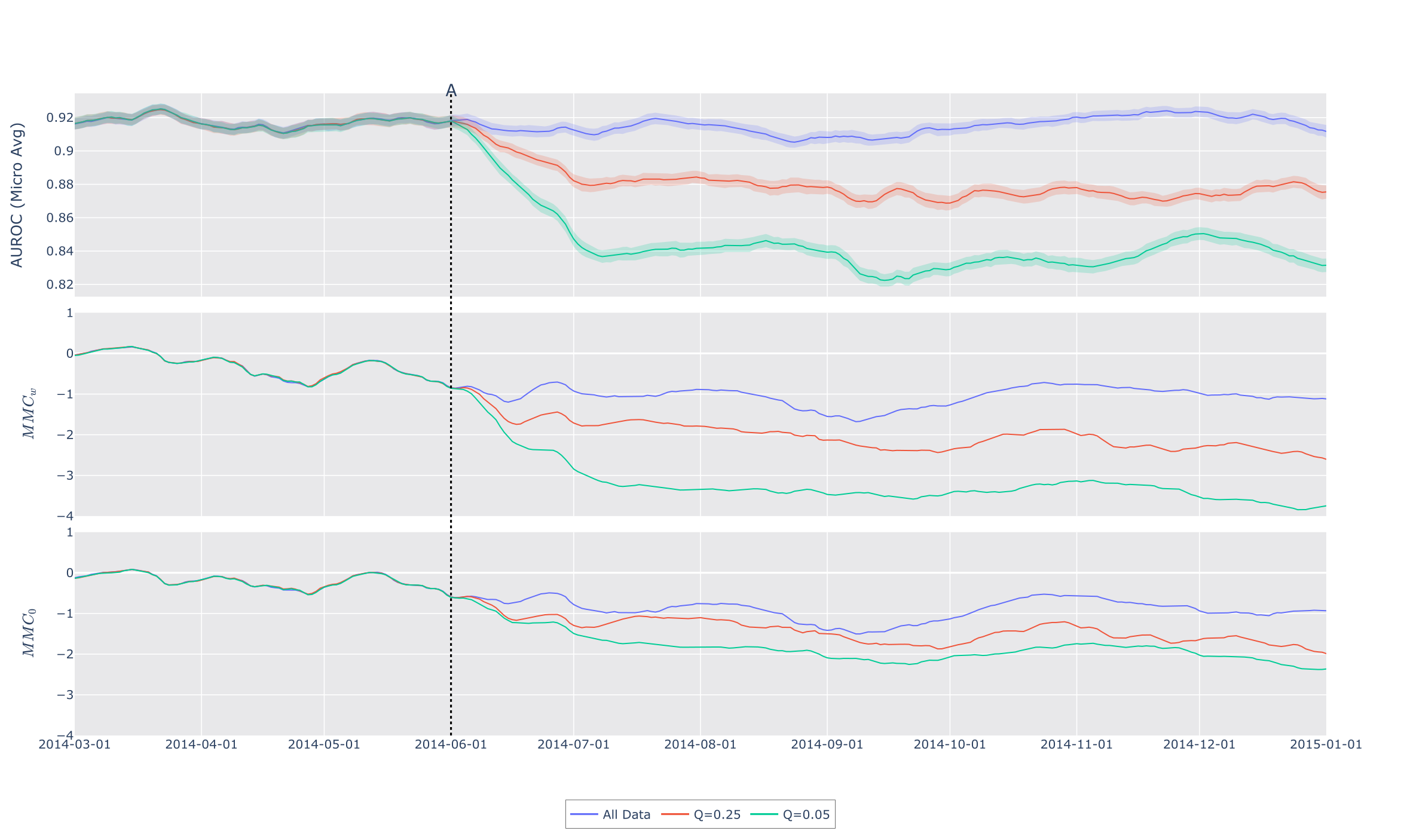}
  \caption{Drift Scenario 1 Results.  Each color represents a different simulated data stream and each panel shows different results: (top) mirco-averaged AUROC across all classes, (middle) weighed multi-modal concordance metric, (bottom) multi-modal concordance metric without any weights (simple average).
  All data streams are the same until point A (2014-06-01), where the composition of the data was modified using hard data mining as described in Section \ref{sec:drift1}.
  Q denotes the quantile value to find predicted probability thresholds for positive and negative samples.  For example, Q=0.25 means the sample pool is formed by the highest 25\% of negative samples and the bottom and lowest 25\% of positive samples (on a per label basis).
  } 
  \label{fig:drift_bad}
\end{figure}

Our principal experiment specifically investigates if performance changes are detectable by our concordance metric.
To accomplish this, we induce performance degradation through hard data mining and observe its effects on concordance and drift.
For this experiment, we mine hard data by including only samples where the classifier has a low degree of certainty on a per label basis by including exams where a pathology was indicated but scored low for that pathology and conversely when no the predicted score was high on an exam but no indication of the pathology was found by an annotator.
Using this method, we created a sample pool that we drew from to populate each day's simulated exams. Every exam during our test set was replaced by an exam randomly drawn from this pool, maintaining the same number of daily exams from the original data stream. See Figure \ref{fig:drift_bad} for performance and MMC plots.

\subsection{Drift Scenario 2: Clinical Workflow Failure}
\label{sec:drift2}

Clinical workflows, especially those that include AI modeling are complex and heavily rely on metadata. There are many situations where this metadata can be inaccurate or inconsistent which cause these workflows to deteriorate or fail over time.  This experiments explores exaggerated cases of these situations in two experiments. The first injects lateral view images which our model was not trained on, and the second adds pediatric data that typical AI systems are not cleared to report on.

\subsubsection{Metadata Filter Failure} 
\label{sec:drift2a}
\begin{figure}[hbt!]
  \centering
  \includegraphics[width=\columnwidth]{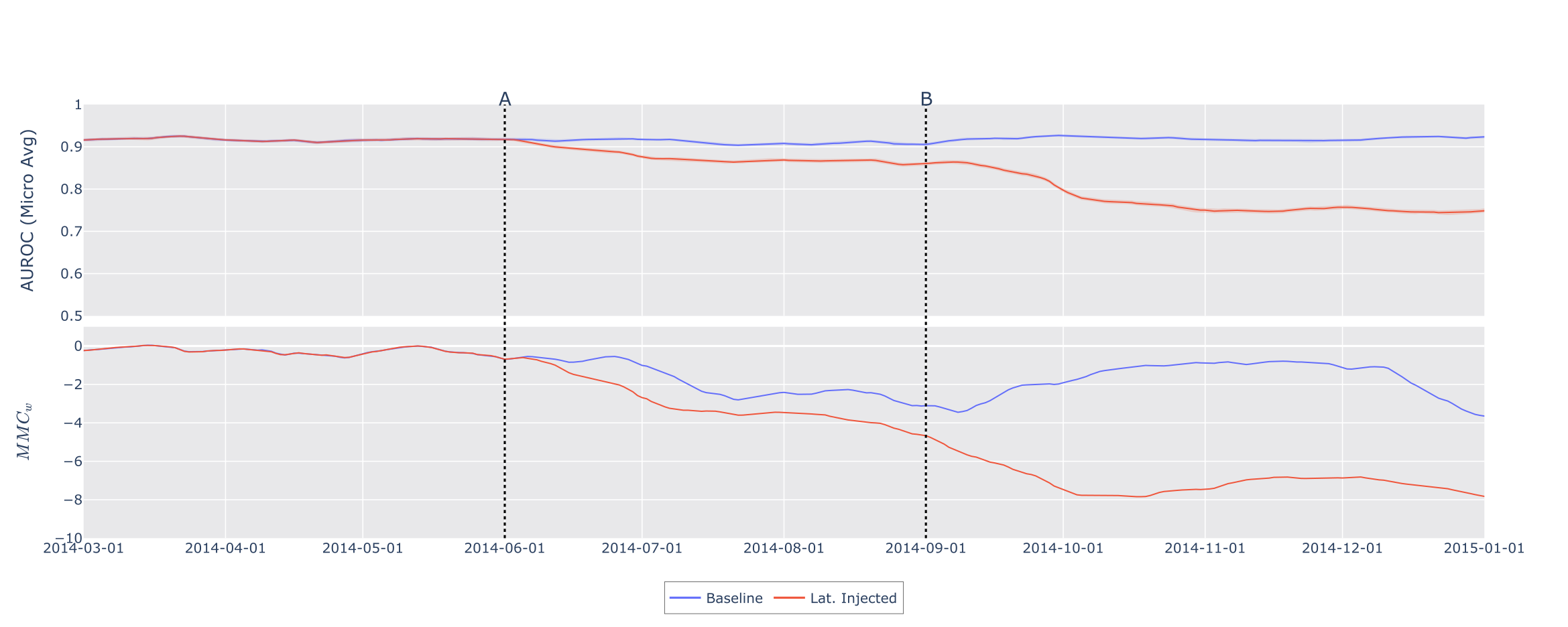}
  \caption{Drift Scenario 2a. Micro-averaged AUROC across all classes on top and our weighed multi-modal concordance metric on bottom
  The baseline (blue) is the unmodified data stream from PadChest. The modified data stream (red) was changed at two points: A) lateral images were added to the stream, B) frontal images were removed from the stream.  See Section \ref{sec:drift2a} for more details.
  }
  \label{fig:drift_lateral}
\end{figure}

For lateral image induced drift experiment, we simulate a failure in a metadata filter by adding lateral images into our test data stream. With the availability of all metadata variables as well as pathology labels on the PadChest lateral images, this use-case enables an end-to-end demonstration of our drift detection pipeline covering each specified metric and input modality. The original PadChest data set includes both frontal and lateral data, so to ''inject`` lateral data, we simply began including these images in the detection windows.  We also simulate a complete failure in this workflow by removing all frontal images, leaving only lateral images. See Figure \ref{fig:drift_lateral} for performance and MMC plots.\par

\subsubsection{No Metadata Available} 
\label{sec:drift2b}
\begin{figure}[htbp!]
  \centering
  \includegraphics[width=\columnwidth]{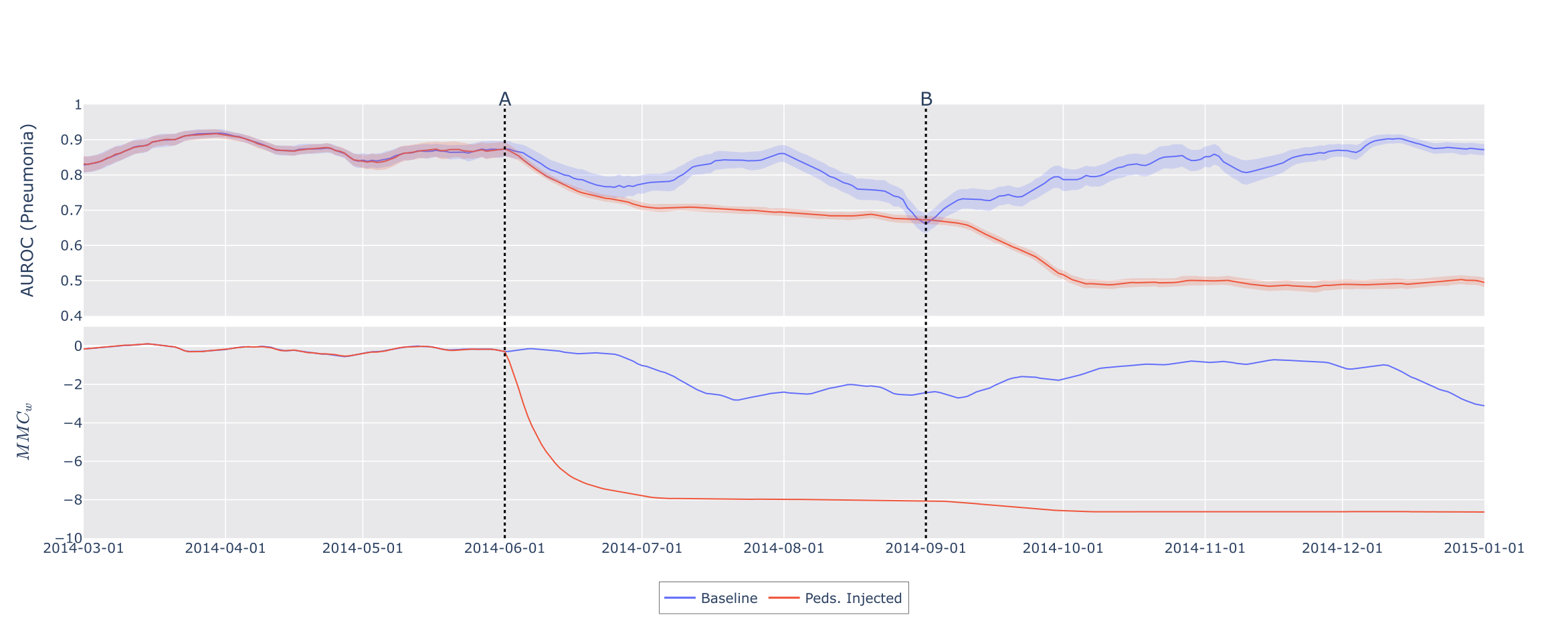}
  \caption{Drift Scenario 2b. Pneumonia AUROC across all classes on top and our weighed multi-modal concordance metric (metadata values removed) on bottom.
  The baseline (blue) is the unmodified data stream from PadChest. The modified data stream (red) was changed at two points: A) images from the Pediatric data were added to the stream, B) PadChest images were removed from the stream.  See Section \ref{sec:drift2b} for more details.} 
  \label{fig:drift_peds}
\end{figure}

This experiment tests how our method works when metadata is missing or unavailable and used the Pediatric Pneumonia Chest X-ray dataset from \cite{Kermany2018LabeledOC}. This dataset contains 5,856 Chest X-rays labeled as either pneumonia or normal and contains only images and labels. Since this dataset has removed all metadata this allows us to investigate how our method performs using only a subset of metrics, particularly the VAE-based drift metrics. As mentioned above, our code repository enables a workflow whereby a user can experiment with their own Chest X-ray dataset, labels, and metadata variables to visualize the entire drift pipeline. The Pediatric Pneumonia Chest X-ray dataset does not include any temporal data, so we injected this data from this dataset similar to the method used in experiment \ref{sec:drift1}, except the entire dataset was used as a sample pool.  As before, the data was sampled randomly without replacement until no samples remained at which point the pool was reset and reshuffled. See Figure \ref{fig:drift_peds} for performance and MMC plots.

\section{Results}
The performance and concordance metrics of each experiment are visualized in Figures \ref{fig:drift_bad}, \ref{fig:drift_lateral} and \ref{fig:drift_peds}. In each figure the top panel depicts performance over time as calculated in each detection window. 
The curves in the second panel of each figure represent our weighted multi-modal concordance metric, $\mmc_w$. In addition, each figure depicts vertical lines which represent points in time that the data stream was modified to induce drift.  Key observations on the three figures are summarized below, where we demonstrate the correlative relationship between our concordance metric and measured performance, illustrating the efficacy of our methodology in clinical AI drift monitoring.

We start by discussing results in Figure \ref{fig:drift_bad} which corresponds to our first experiment outlined in Section \ref{sec:drift1}.
In this figure, the top panel shows micro-averaged AUROC, the middle depicts our metric $\mmc_w$, and the bottom panel shows our metric without the use of performance correlative weights ($\mmc_0$).
As mentioned in Section \ref{sec:drift1}, this experiment induces performance drift by limiting samples to hard data.  In the figure legend the Q values refers to the degree to which the samples were limited.
Specifically, Q denotes the quantile of predicted probabilities used to find high scoring negatives and low scoring positives, i.e $Q=0.25$ mean that the highest 25\% of negative and lowest  25\% of positive samples (by predicted probability values) were used. ''All data`` set this quantile value to $1.0$, thereby using all the negative and positive samples.
In this figure, we see clear correlation between measured ground truth and our concordance metric when comparing the top panel (performance) to both the middle and bottom panels ($\mmc_w$ and $\mmc_0$ respectively).
First, we notice that the baseline does not show a visible drop. This is representative of a clinical scenario where an exceedingly well performing model remains as effective as baseline (validation) over the course of production. 
We also see that as we reduce the Q value, thereby decreasing performance, we see a corresponding drop to the concordance metric.
In this experiment, we also depict both the weighted and unweighted version of our metric and we can see the value of using performance correlated weights when comparing the middle and bottom panels.  When using the weighted version of our multi-modal metric, we see very clear separation of all 3 curves correlative to the performance changes in the graph.  However, in the bottom panel, we see the 3 curves closely clustered.  This comparison shows how our weighting methodology emphasizes relevant metrics to yield a consistent performance proxy.

Figure \ref{fig:drift_lateral} depicts the results of our second experiment that investigates clinical workflow failures that could lead to model drift, described in Section \ref{sec:drift2a}. In this figure, again we show performance in the top panel and our metric result in bottom panel. In this experiment, we have two trials, a baseline which is the original data stream of PadChest (blue) and a second trial (red) drift has been induced. We show two vertical lines which denote the point in time where we modify the data stream in the second trial.  At point A, we simulate a failing metadata workflow by allowing lateral images to be passed to the model, and at point B we simulate a catastrophic failure in the metadata workflow by removing in-distribution (frontal) data, leaving only laterals.  
In this experiment, we again see a correlative response in $\mmc_w$ to performance as shown in Figure \ref{fig:drift_lateral}.  At point A, where we introduce lateral images, both performance and $\mmc_w$ drop a modest amount, with AUROC dropping from above $0.9$ to around $0.85$ and $\mmc_w$ dropping to about $-4$. At point B, we see an another decrease in performance (to $\approx0.75$ AUROC) and our metric drops to, and hovers around $-8$.  This demonstrates our method's robustness to changes in data composition that are detectable by metadata tags as well as visual appearance.

Results for our final experiment appear in Figure \ref{fig:drift_peds}, which simulates a situation where metadata is unavailable or is not used properly in a clinical workflow and the model begins to receive images that it is has not been validated to classify. This experiment, described in Section \ref{sec:drift2b}, investigates how well our system performs without the use of any metadata, relying only on VAE latent representations and predicted probability distribution shifts to detect drift.
The performance metric in this experiment differs slightly from the other two in that the we only measure Pneumonia AUROC as the pediatric data includes only pneumonia labels, invalidating performance measures for other classes. At point A in this experiment, we modify the data stream by injecting pediatric cases such that for every exam in the original PadChest data stream there are 3 pediatric exams added.  Then at point B, the data stream switches over to only pediatric data.  
As seen in Figure \ref{fig:drift_peds},  
unsurprisingly, the performance metric fluctuates more than in the other figures as this represents performance of a single pathology rather than an average.
Comparing the baseline performance and that of the trial with a modified data stream, we see a drop in performance and concordance at both points where we modify the data stream. Even though the baseline performance oscillates between $\approx0.9$ and $\approx0.7$, when we inject pediatric data the AUROC drops and remains consistently below the baseline performance.
Likewise we see a significant drop in $\mmc_w$ at both data stream modification points.
This indicates that our approach remains robust even when metadata is not available and can successfully detect drift using only VAE latent representations and predicted probabilities.
In this experiment, we also notice an exaggerated drop in $\mmc_w$ at point A compared to performance. This large drop in concordance is expected since pediatric data is, indeed, out-of-distribution, but we likely see a smaller drop in performance since the specific features of pneumonia in these images are similar enough to those in frontal PadChest images that the classifier can pick them up, therefore performance is less affected by this drift.
Concordance takes a more holistic approach to data drift and these images represent significant drift in the data stream; so in this case, the drift metrics are more sensitive to data stream changes even though the classifier is more robust to those same changes.  This is a desirable characteristic for concordance measurement since AI models are not typically cleared for use on pediatric patients, so when the system is flooded with patients under 12 - alerting to data drift is appropriate, regardless of performance changes.

\section{Discussion}

The purpose of this work was to explore a data-driven approach to building a system that can perform real-time AI model monitoring for a medical imaging model over time and identify methodologies to achieve useful drift/performance metrics in the absence of contemporaneous ground truth in a chest X-ray model use case. We found that, rather than measuring change in per-image reconstruction loss which is a scalar quantity, utilizing the $z$ vector (latent space encoding generated by the VAE) provides a feature-rich representation with compressed yet relevant information about input X-rays. Furthermore, we found that we were able to generate a strong proxy for ground truth performance using this latent representation along with relevant DICOM metadata tags and distributional shifts between predictive model probabilities.  By unifying concordance metrics captured from this data, we present our multi-modal approach that can monitor real time medical imaging AI systems. We demonstrate through experimentation that this approach to unsupervised drift detection correlates with supervised performance drift and has crucial implications on addressing the translation gap related between continuous model performance modeling in dynamic healthcare environments that lack contemporaneous ground truth.  

When we monitor drift, one objective is to inform decisions regarding the model performance in production with the expectation that if data distributions are similar between training and production then the model should perform as expected. If the distributions have changed, the whole system might need an update.The task of drift detection focuses on global data distributions in the whole dataset in order to determine if there is significant shift compared to the past period data or model training data.   Data drift might occur as a gradual shift in features along one of many potential dimensions; the relationship to model performance will determine the need for intervention. This is different conceptually from the traditional task of out of distribution detection where the focus is to find individual "unusual" or "different" features for input data. In other words, the global data drift and out of distribution outliers can exist independently; the entire dataset might drift without outliers just as an individual outlier might appear without data drift. If drift is detected in this framework the goal is to intervene at the model level (i.e. pull it out of production, retrain, rebuild, etc.). In contrast if out of distribution input is detected the assumption is made that the model still performs well but for that particular input data the prediction would not be accurate and intervention would occur at the data level (i.e. logic is applied to qualify the model decision or avoid model). Safe and effective monitoring in production requires that both global drift and out-of-distribution detection can be identified with different interventions needed, and in the medical imaging use case, that is further challenged in that ground truth is not immediately available.  In this framework, model monitoring must accommodate both the gradual global drift as well as the individual case of an outlier; in the former case this work aimed to design for the medical imaging use case as a drift detection metric to alert for potential conditions that would impact overall model performance in the absence of ground truth. 

Current proposed workarounds for the lack of ongoing medical imaging AI model performance monitoring solutions include primarily relying on human experts/users to provide model feedback during deployment and/or perform periodic expert auditing on retrospective data by curating representative data and applying ground truth labels for performance analysis. This is an unsustainable and problematic solution for several reasons. First, asking end users to add additional cognitive effort (and clicks) in order to provide model feedback risks decreasing the purported efficiency advantages of using AI models adding to user burnout. Second, while it is generally agreed upon that periodic "model audits" that resemble the initial pre-deployment analysis will be important to ensure ongoing model performance in production, the institutional effort involved in curating, annotating, and analyzing model performance are powerful disincentives. These approaches are difficult to scale in an environment with limited resources and, potentially, many different models deployed in a given clinical practice.  One needs to decide when to perform these checks, on what data and how frequently, and leverage expensive expert clinical resources to repeatedly curate and annotate ground truth test datasets. Therein lies the critical trade-off between maintaining consistent patient care and burdening clinicians to evaluate performance, a burden compounded by the fact that individual annotation (with inherent variance) causes noisy feedback. Worse, drift can occur across innumerable axes and features across the imaging domain, which risks overlooking important changes to performance leading to patient harm. This approach not only provides insufficient performance details, it fails to address medical imaging AI systems that do not directly interact with the end user in real time (i.e. image reconstruction, autonomous screening workflows, etc), or perform super-human tasks (i.e. opportunistic screening, mortality prediction), and further, risks missing unconscious biases. To combat the divergence between static models and their dynamic clinical environments, strategies to detect drift in real time must be developed and adopted.  We present a multi-modal and sequential drift detection system for medical image classifiers, which can be modified flexibly to fit different data domains. Previous work has mainly been limited to a certain type of data, like streaming text \cite{jl1}, image and video \cite{jl2} or metadata-like informational markers from clinics, airlines, internet of things (IoT), etc. \cite{jl3}. Even though their methods could be expanded into other data types, multiple metrics for drift detection would be generated. Integrated analysis of clinical data and medical images (pathology and radiology) is routine in clinical practice and the lack of multi-modal models which perform such integration represents a significant gap. Our system generates a unified metric from multiple features using standardizing and weighting strategies, which provides a more holistic evaluation to aid in decision-making.

In a production environment for a medical imaging AI model, drift metrics are likely to serve as the primary signal that model performance might have changed in the absence of real-time access to ground truth; we have demonstrated that statistical change in model inputs and outputs may serve as a valuable proxy that can signify a possible decay in the model performance. Ideally, a model experiencing significant drift would be paused from production, allowing for confirmatory performance auditing, identifying potential root causes, intervening with using a new model, current model retraining, refactoring DICOM study routing workflows, and ideally, inform continuous learning strategies. But the actions available to respond to drift in real-time production in medical imaging are constrained by common regulatory environments. For example, the current model approval process by the FDA may discourage continuous model code changes and updates, as they may trigger a re-submission for approval. This in turn discourages continuous monitoring and reporting of flaws due to the high overhead required to provide updates to mitigate these performance gaps. This regulatory challenge does not, however, change the fundamental need for model monitoring to achieve safe, effective model deployment and that the detection of model drift is not itself sufficient without a mechanism to address the disparity in a timely manner.  In the future, as expansion of regulatory permissions are expected to include retraining/continuous learning informed by Machine Learning Operations (MLOps) monitoring systems such as the one proposed, the information provided by drift metrics could inform a learning healthcare system allowing for intelligent model monitoring, auditing, retraining, and redeployment \cite{pianykh2020continuous, Finlayson2021TheCA}. 

This work has several important limitations. First was the use of public datasets that precluded access to additional clinical contextual information and important population metrics.  Importantly, the primary dataset was chosen due to the preserved temporal relationships between studies, availability of DICOM metadata, and well-labeled ground truth for pathologies in order to simulate a distribution of imaging examinations over time.  While useful for temporal relationships, the dataset was highly curated retrospectively and the sustained performance of the baseline model over time likely is a reflection of that curation homogeneity rather than the lack of underlying model drift; a real world sequential dataset with ground truth labels would serve as a more realistic baseline with which to observe and detect routine drift over time in the absence of manufactured drift experiments as was done in this work. There are many other important use cases related to medical imaging not explored including incorporating new disease data, impact of hospital protocol or equipment data, and more. Further the complexity of medical imaging use cases are varied and include increasing data size and complexity (i.e. MRI reconstruction, CT imaging, etc) as well as varied clinical tasks (i.e. segmentation, diagnosis, outcome prediction) which were not explicitly evaluated in this work. Nonetheless, the underlying methodological approach to medical imaging model monitoring explored and validated by this work could be further investigated in these and other important use cases.  

In conclusion, this work demonstrated a system that can perform AI model monitoring for a medical imaging with methodologies that can achieve real-time drift metrics in the absence of contemporaneous ground truth in a chest X-ray model use case to inform potential change in model performance. This work will inform further development of automated medical imaging AI monitoring tools to ensure ongoing safety and quality in production to enable safe and effective AI adoption in medical practice. The important contributions include the use of VAE in reconstructing medical images for the purpose of detecting input data changes in the absence of ground truth labels, data-driven unsupervised drift detection statistical metrics that correlate with supervised drift detection approaches and ground truth performance, and open source code and datasets to optimize validation and reproducibility for the broader community.

\section*{Acknowledgments}
This work was was supported in part by the Stanford Center for Artificial Intelligence in Medicine and Imaging (AIMI) and Microsoft Health and Life Sciences.

\bibliographystyle{unsrt}  
\bibliography{references}  

\clearpage
\appendix
\section*{Appendices}

\section{Training Configuration}
\label{sec:model-train}

We release the full code to train, validate and test our approach. The mapping for each label from PadChest radiological findings appears in Table \ref{tab:labels}. 
Please see our code at \coderepourl for more training details.

\begin{table}[hp]
\caption{Unified CheXpert and PadChest labels Mapping. New labels in bold on the left and radiological finding from PadChest reports on the right}
\vspace{1em}
\label{tab:labels}
\makebox[\textwidth][c]{%
    \begin{minipage}{.95\textwidth}
        \begin{description}
            \item[Atelectasis] \hfill laminar atelectasis, fibrotic band, atelectasis, lobar atelectasis, segmental atelectasis, atelectasis basal, total atelectasis
            \item[Cardiomegaly] \hfill cardiomegaly, pericardial effusion
            \item[Consolidation] \hfill consolidation
            \item[Edema] \hfill kerley lines
            \item[Lesion] \hfill nodule, pulmonary mass, lung metastasis, multiple nodules, mass
            \item[No Finding] \hfill normal
            \item[Opacity] \hfill infiltrates, alveolar pattern, pneumonia, interstitial pattern, increased density, consolidation, bronchovascular markings, pulmonary edema, pulmonary fibrosis, tuberculosis sequelae, cavitation, reticular interstitial pattern, ground glass pattern, atypical pneumonia, post radiotherapy changes, reticulonodular interstitial pattern, tuberculosis, miliary opacities
            \item[Pleural Abnormalities] \hfill costophrenic angle blunting, pleural effusion, pleural thickening, calcified pleural thickening, calcified pleural plaques, loculated pleural effusion, loculated fissural effusion, asbestosis signs, hydropneumothorax, pleural plaques
            \item[Pleural Effusion] \hfill pleural effusion
            \item[Pneumonia] \hfill pneumonia
        \end{description}
    \end{minipage}%
}
\end{table}

\subsection{Training Hyperparameters}
\begin{table}
\centering
\caption{Training Hyperparameters}
\label{tab:model-train}
\vspace{1em}
\begin{subtable}[t]{0.49\textwidth}
\label{tab:class-train}
\centering
    \begin{tabular}{llr}
\toprule
{} & CheXpert &   PadChest \\
\midrule
Learning Rate &        0.0001 &    0.001 \\
Max Epochs    &        50 &   48 \\
Step Size     &        10 &   12 \\
Gamma         &        0.9 &    0.1 \\
Weight Decay  &        0.00001 &    0.00001 \\
Image Size    &        320x320 &    320x320 \\
\bottomrule
\end{tabular}
\subcaption{Multi-label Classifier Training Hyper-parameters}
\end{subtable}
\hfill
\begin{subtable}[t]{0.49\textwidth}
\label{tab:vae-train}
\centering
\begin{tabular}{rll}
\toprule
&  Parameter &  Value \\
\midrule
\multirow{6}{*}{\rotatebox{90}{Optimization}} & Base Lr &         0.0001 \\
    & Batch Size &            128 \\
    & Max Epochs &             50 \\
    & Lr Scheduler &        ReduceOnPlateau \\
    & Gamma &            0.1 \\
    & Cooldown &              0 \\ \midrule
\multirow{6}{*}{\rotatebox{90}{VAE}} & Image Dims &  [1, 128, 128] \\
    & Normalize &          False \\
    & Layer Count &              4 \\
    & Width &            240 \\
    & Z &            128 \\
    & KL Coeff &            0.1 \\
\bottomrule
\end{tabular}
\subcaption{VAE Training Hyper-parameters}
\end{subtable}
\end{table}

\section{Supplemental Results}

\subsection{VAE Results}

We outline key results pertaining to the VAE-based drift detector, and visualize the detector’s ability to identify drifted data through multiple signals.
We find that leveraging the latent space representation as opposed to the scalar reconstruction loss yields a richer encoded feature set within which to contextualize the VAE’s ability to separate out-of-distribution data in production. The VAE’s encoded latent space representation ($z$) consists of a 128-length vector. To demonstrate that the VAE drift detector is contextually meaningful, we present indices of the latent encoding that exhibit strong correlations with drifted inputs (metadata features representing opposing categories, low to high activation scores, lateral X-ray images etc.). 

\begin{figure}[!htp]
  \centering
  \includegraphics[width=\columnwidth]{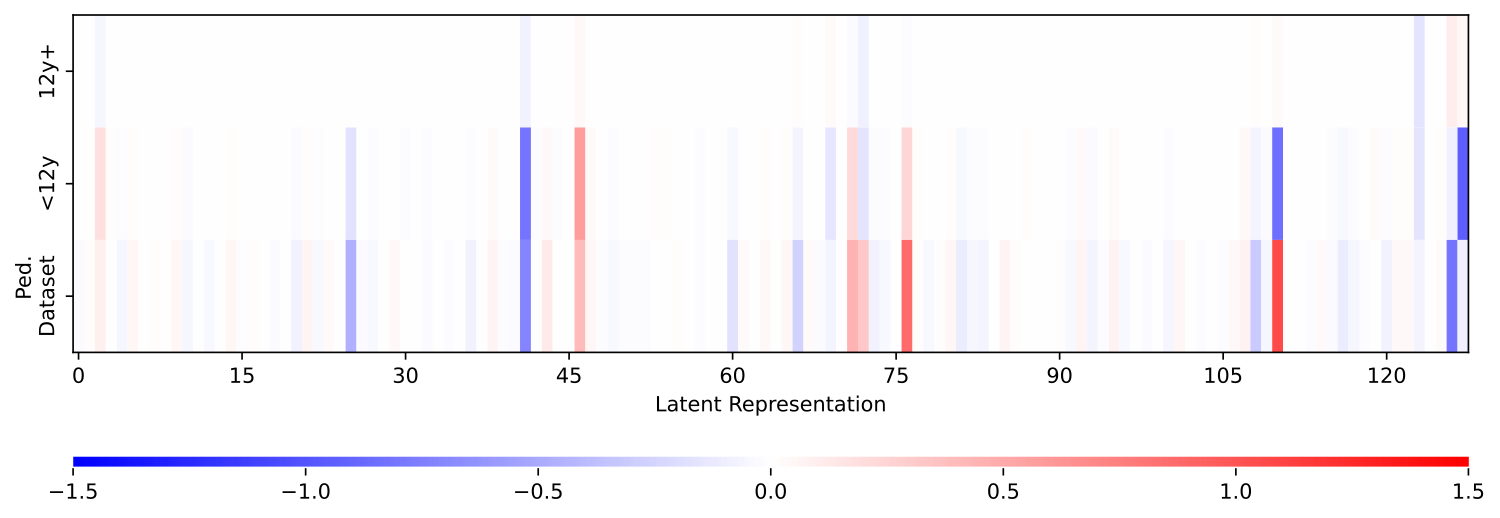}
  \caption{Averaged Latent Representations of X-Ray Images Split by Patient Age. (top) PadChest dataset, patients over 12 years old, (middle) PadChest dataset, patients 12 years old and under, (bottom) Images from the Pediatric Dataset.}
  \label{fig:vae_peds}
\end{figure}

\begin{figure}[htp]
  \centering
  \includegraphics[width=\columnwidth]{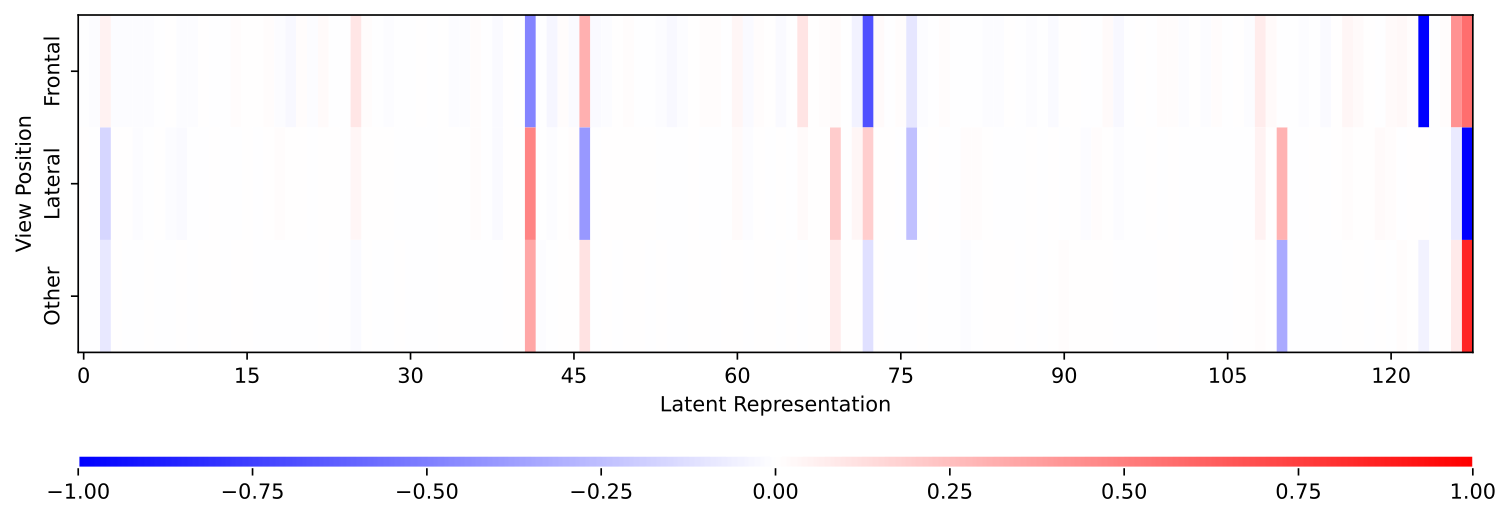}
  \caption{Averaged Latent Representations of X-Ray Images Split by View Position.  (top) frontal view positions, PA, AP, and AP\_horizontal, 
  (middle) lateral view positions, LATERAL, LL, RL, LLD, (bottom) any view position not included in frontal or lateral images.
  }
  \label{fig:vae_view}
\end{figure}

\begin{figure}[htp]
  \centering
  \includegraphics[width=\columnwidth]{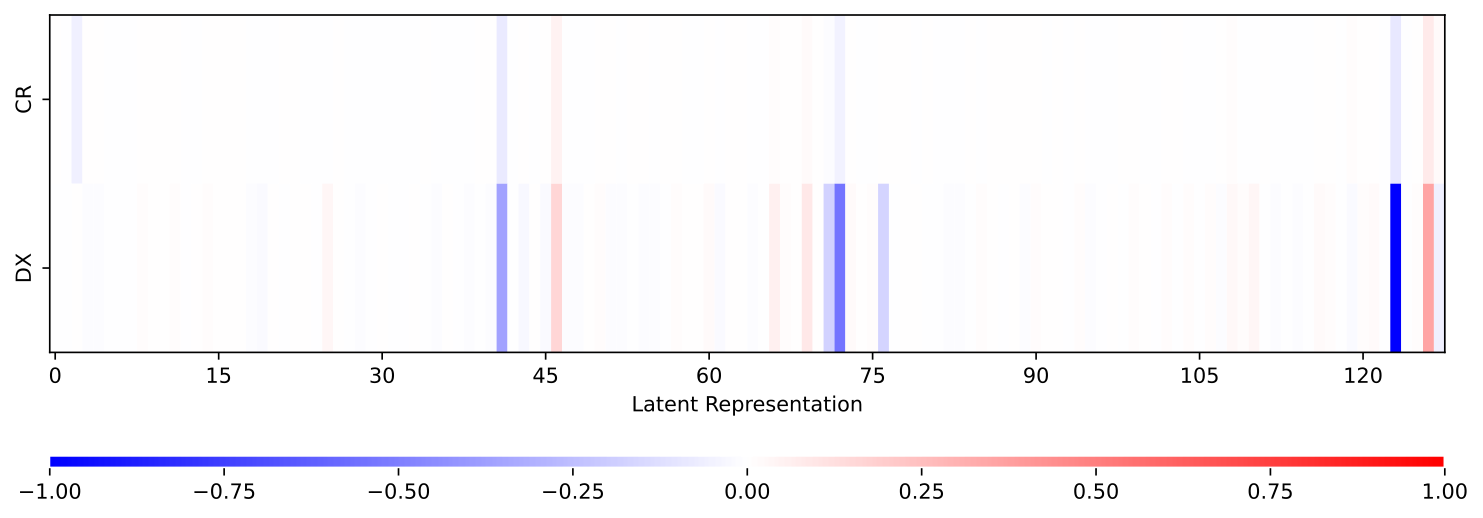}
  \caption{Averaged Latent Representations of X-Ray Images Split by Imaging Modality. (top) CR, computed radiography exams, (bottom) DX, digital x-ray exams}
  \label{fig:vae_modality}
\end{figure}

Figures \ref{fig:vae_peds}, \ref{fig:vae_view}, \ref{fig:vae_modality} show VAE latent representations of images averaged over different populations or image properties.  These figures show how (on average) VAE will encode images with different characteristics.  Within all of these figures we see trends within individual latent values, which enable our multi-modal concordance metric to capture detect drift in the appearances of images. Here we highlight changes that specifically related patient population and image construction as an example.
Figure \ref{fig:vae_peds} shows average latent representations of PadChest exams of patients over 12 years old, 12 years old and under, as well as averaged latent representations from the Pediatric dataset.
We see significant differences in the latent space of x-rays of older patients and patients 12 and under.  In the pediatric data, we see a similar representation to images of patients under 12 from the PadChest dataset, with some differences likely attributed other differences in image formation from the two datasets.
Figure \ref{fig:vae_view} shows the average latent space representations of frontal, lateral and other view positions.  Again we see significant differences in their average encoded spaces.  In particular, we see a number of values diametrically opposed between the frontal and lateral vectors, demonstrating the large differences between these view positions
In Figure \ref{fig:vae_modality}, we show averaged latent spaces for each modality as reported by the DICOM metadata. The majority of the data is CR, so we expect a fairly uniform average as expected.  With DX images, we see certain latent variables exaggerated as compared to CR images.

\end{document}